\documentclass[11pt,twoside]{article}

\usepackage{asp2006}
\usepackage{graphicx}

\begin{document}
\title{The stars and gas in outer parts of galaxy disks: 
Extended or truncated -- flat or warped?}   
\author{P.C. van der Kruit}   
\affil{Kapteyn Astronomical Institute, University of Groningen, P.O. Box 800,
9700 AV Groningen, the Netherlands; 
vdkruit@astro.rug.nl}    

\begin{abstract} 
I review observations of truncations of stellar disks and models for 
their origin, compare observations of truncations in 
moderately inclined galaxies to those in edge-on systems and discuss the
relation between truncations and HI-warps and their systematics and 
origin. Truncations are a common feature in edge-on stellar disks, but the 
relation of truncations in face-on to those in  edge-on galaxies needs further 
clarification. The origin of truncations is most likely related to a maximum
in the specific angular momentum in the material that formed the stellar
disks, but this model 
does probably require some redistribution of angular momentum.
HI-warps start just beyond the truncation radius and disks and warps appear
distinct components. This suggests that inner disks form initially and 
settle as rigid, very flat structures, while HI-warps result from later infall
of gas with a different orientation of the angular momentum The
\LaTeX-Beamer  presentation of this review is available in pdf-format at
www.astro.rug.nl/$\sim $vdkruit/jea3/homepage/vaticanpres.pdf.
\end{abstract}



\section{Truncations in edge-on galaxies}

Truncations were first found in edge-on spiral galaxies, where the 
remarkable feature was noted that the radial 
extent did not grow with deeper photographic exposures 
\citep{vdK79}. Especially, 
when a bulge was present the minor axis did grow considerably 
on IIIA-J images compared to the Palomar Sky Survey IIa-D exposures in 
contrast with the major axes.
Prime examples of this phenomenon of ``truncations'' 
(originally also called ``cut-offs'') are the edge-on galaxies NGC
4565 and NGC 5907 (Fig. 1). The truncations appear very sharp,
although of course not infinitely.
\Citet{vdKS81} state:
{\it ``This cut-off is very sharp with an e-folding of less than about
1 kpc'', based on the spacing of the outer isophotes.''}
Sharp outer profiles are actually obtained
after deprojecting near-IR observations of edge-on galaxies
\citep[e.g.][]{Flo06}.

Various models have been proposed for the origin of truncations:\\
{\bf I}.
The truncations are the current extent of disks that are growing from the 
inside out from accretion of external material \citep{Lar76}.
This predicts substantial age gradients across disks, which are not
observed \citep{deJ96b}.
Current thinking is that disks formed in an initial monolythic collapse 
followed by a protracted period of infall of gas and capture of dwarfs 
companions or 
by a slow continuing process of merging of existing systems in a 
hierarchical picture.\\
\begin{figure}[!ht]
\includegraphics[width=13cm]{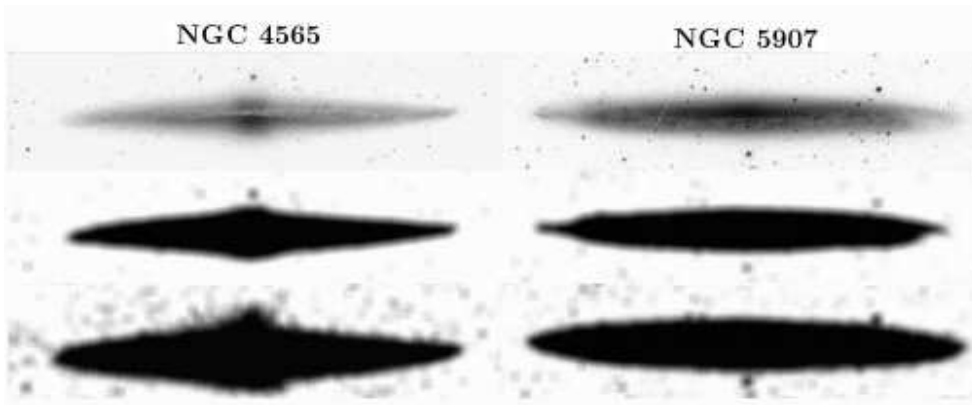}
\caption{NGC4565 and NGC 5907 at various light levels (from van der
Kruit (2007).}
\end{figure}
{\bf II}.
Inhibition of star formation when the gas surface (or space?) density 
falls below a threshold for local stability \citep{FE80,Ken89,Sch04}.
The Goldreich-Lynden-Bell criterion for stability of gas layers
gives a poor prediction for the truncation radii \citep{vdKS82}.
Another problem is that the rotation curves of some galaxies --e.g. NGC 5907 
and NGC 4013 \citep{Cas83,Bot96}-- show features
near the truncations that indicate that the {\bf mass}
distributions are also truncated.
Schaye predicts an anti-correlation between $R_{\rm max}/h$ and $h$,
which is not observed.\\
{\bf III}.
The truncation corresponds to a maximum in the specific angular momentum 
distribution in the protogalaxy \citep{vdK87}. If the collapse occurs from 
a  \citet{Mes63} sphere (i.e. uniform density and 
angular rotation)  with detailed conservation of specific
angular momentum in the force field of a dark
halo with a flat rotation curve, a roughly exponential disk
results with a truncation at about 4.5 scalelengths.
This provides at the same time an explanation for the exponential nature of
disk as well as for the occurence of the truncations.\\
{\bf IV}. It is also possible that substantial redistribution of angular 
momentum, takes place, such that its distribution now is  
unrelated to the initial distribution in the material that formed the disks.
Bars may play an important role in this, as suggested by \citet{Erw07}.
In fact a range of possible agents, such as bars, density waves,
heating and stripping of stars by bombardment of dark matter subhalos, has
been invoked \citep{deJ07}.\\
{\bf V}.
The magnetic model \citep{Bat02,Flo06}, in which a magnetic force breaks
down as a result of star formation so that stars escape.
The evidence for sufficiently strong magnetic fields
needs  strengthening.

\begin{figure}[!ht]
\includegraphics[width=13cm]{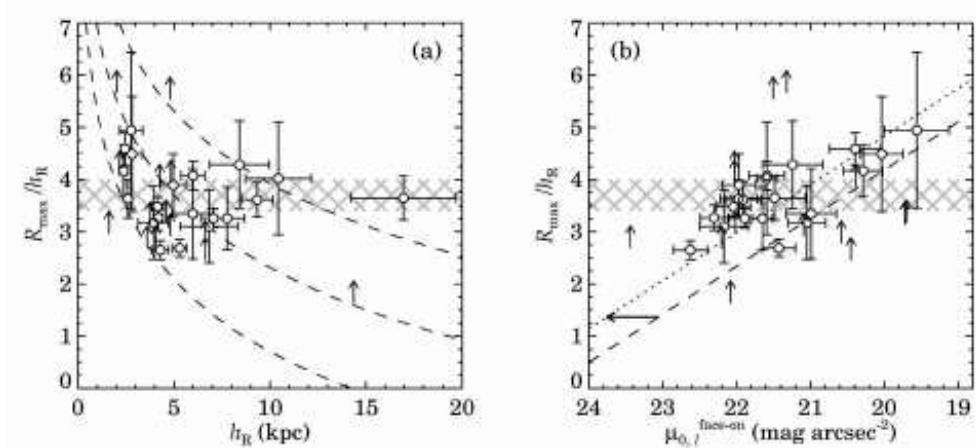}
\caption{Correlations of $R_{\rm max}/h$ with scalelength $h$ and 
face-on central surface brightness $\mu _{\circ, fo}$
(from Kregel \&\ van der Kruit 2006). The dotted lines show predictions from
the star formation theshold model of Schaye (2004).}
\end{figure}

\Citet{KvdK04} derive correlations of the ratio of the cut-off radius in 
terms of disk scalelengths with $h$ itself and with the face-on central surface
brightness $\mu _{\circ, fo}$ (Fig. 2).
$R_{\rm max}/h$ does not depend strongly on $h$, but is somewhat less 
than the 4.5 predicted from the collapse from a simple Mestel-sphere.
There is some correlation between $R_{\rm max}/h$ and $\mu _{\circ, fo}$, 
indicating approximate constant disk surface density at the truncations,
as possibly expected by the star-formation theshold model.
But this model predicts an anti-correlation between $R_{\rm max}/h$ and 
$h$ \citep{Sch04}, which is not observed (see Fig. 2a).
The maximum angular momentum hypothesis predicts that
$R_{\rm max}/h$ should not depend on $h$ or $\mu _{\circ, fo}$ and 
such a model therefore requires some redistribution of angular momentum in the 
collapse or somewhat different intial conditions.

From these results  concerning edge-on galaxies I conclude:
{\bf (1)} Many, but not all, stellar disks
in  edge-on galaxies show evidence for relatively
sharp truncations in their radial distributions.
{\bf (2)}  The model with a theshold in star formation as the origin of the
truncations \citep{Sch04} is not in agreement with the observed distribution 
of $R_{\rm max}/h$  with $h$. 
{\bf (3)} If the truncation radius corresponds to a maximum in the specific 
angular momentum that existed already before the collapse and 
is conserved through the collapse, the initial configuration is
either not identical to that of a uniform density, uniform angular momentum
Mestel sphere and/or the conservation of specific angular momentum is not
perfect.

\section{Truncations in moderately inclined galaxies}

Due to line-of-sight integration truncations should more difficult to
detect in face-on galaxies than in edge-on ones.
The expected surface brightness at 4 scalelengths is
about 26 B-mag arcsec$^{-2}$ or close to sky.
In face-on galaxies like
NGC 628 \citep{Sho84,vdK88} an isophote map shows that 
the outer contours have a much smaller spacing than the inner ones.
The usual analysis uses an inclination and major axis determined from
kinematics (if available, otherwise this is estimated from the
average shape of isophotes) and then determines an azimuthally averaged
radial surface brightness profile. But this will smooth out any truncation
if its radius is not exactly constant with azimuthal angle. 
\begin{table}[!ht]
\caption{Studies of truncations in galaxy disks}
\smallskip
\begin{center}
\begin{tabular}{||l|c|c|c|c||}
\hline
authors & orient. & trunc. & no trunc. & up-bend. \\
\hline
\citet{vdKS82} & edge-on & 8 & -- & -- \\
\citet{Poh00} & edge-on & 30 &  -- & -- \\
\citet{Poh02} & face-on & 3 & -- & -- \\
\citet{KvdK04} & edge-on & 20 & 11 & -- \\
\citet{PT06} & face-on & 54 & 9 & 21 \\
\citet{vdK07} & edge-on & 19 & 7 & -- \\
\citet{Poh07} & edge-on & 9 & 1 & 1 \\
\hline
\end{tabular}
\end{center}
\end{table}
\begin{figure}[!ht]
\includegraphics[width=6.5cm]{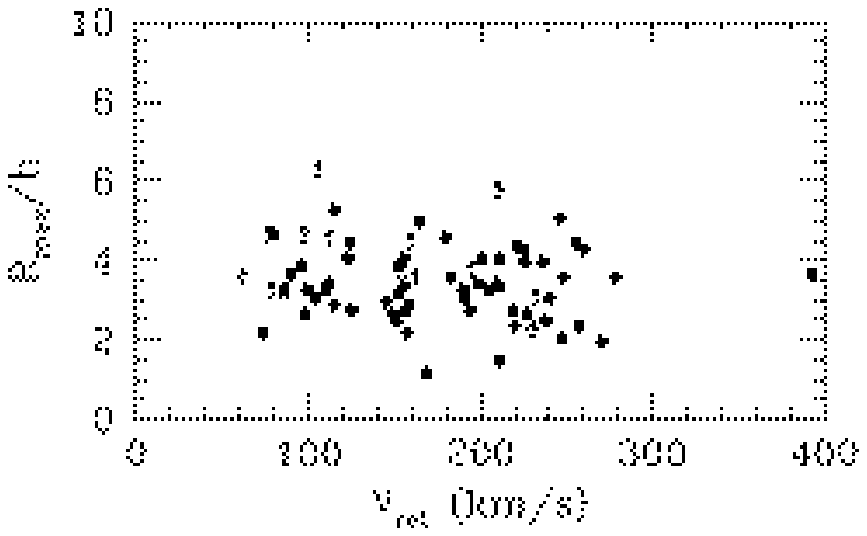}
\includegraphics[width=6.5cm]{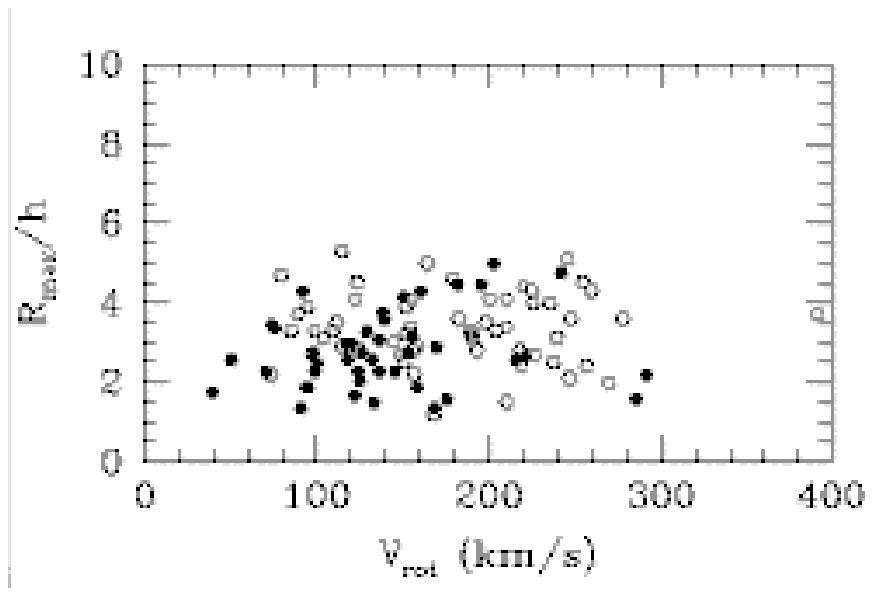}
\includegraphics[width=6.5cm]{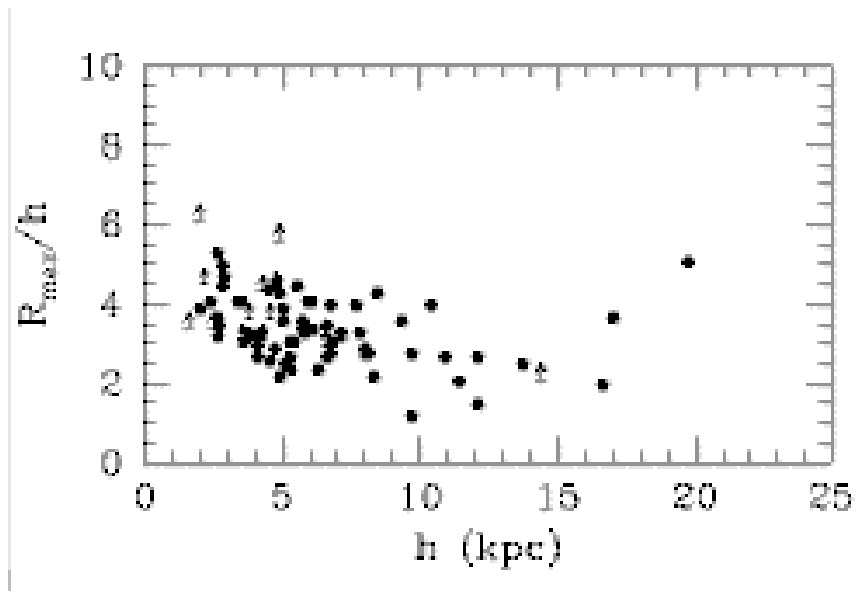}
\includegraphics[width=6.5cm]{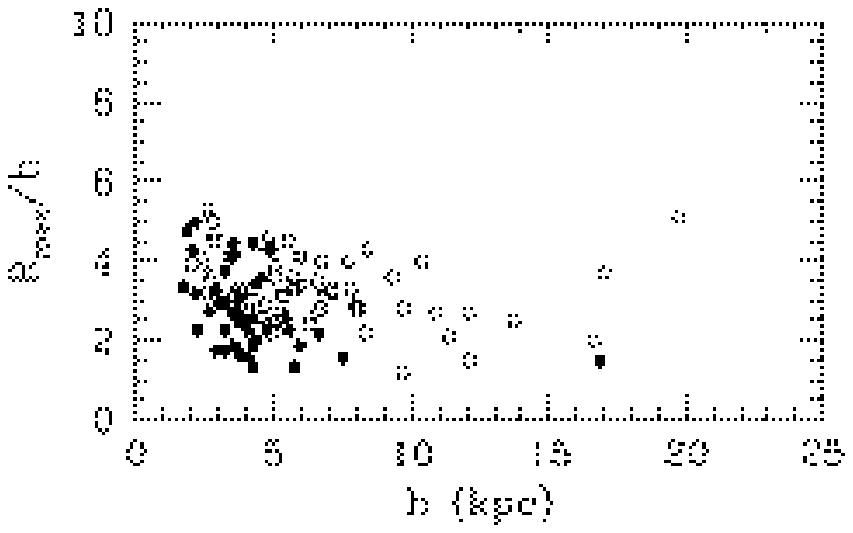}
\caption{Plots of $R_{\rm max}/h$ versus $V_{\rm rot}$. The panel top-left 
has only edge-on systems; in the panel top-right 
these are open circles and filled dots 
show the moderately inclined systems of Type II (''down-bending break'').
The lower panels show the same for $R_{\rm max}/h$ versus $h$.}
\end{figure}
The effects are nicely illustrated in the study of NGC 5923
by \citet{Poh02} (their Fig. 9), which has isophotes in polar coordinates.
The irregular outline shows that some smoothing out will occur contrary
to observations in edge-on systems.
Unless special care is taken we will always find a (much) less sharp truncation 
in face-on than in edge-on systems.

Table 1 summarizes the samples in the optical, which have been studied for
the presence of truncations.
For almost all of these galaxies the rotation velocity can be found using
HYPERLEDA.
\Citet{PT06} studied a sample of moderately inclined systems 
through ellipse-fitting of isophotes in SDSS data.
They distinguish three types of profiles:
{\it Type I}: no break;
{\it Type II}: downbending break;
{\it Type III}: upbending break.
\begin{figure}[!ht]
\includegraphics[width=6.5cm]{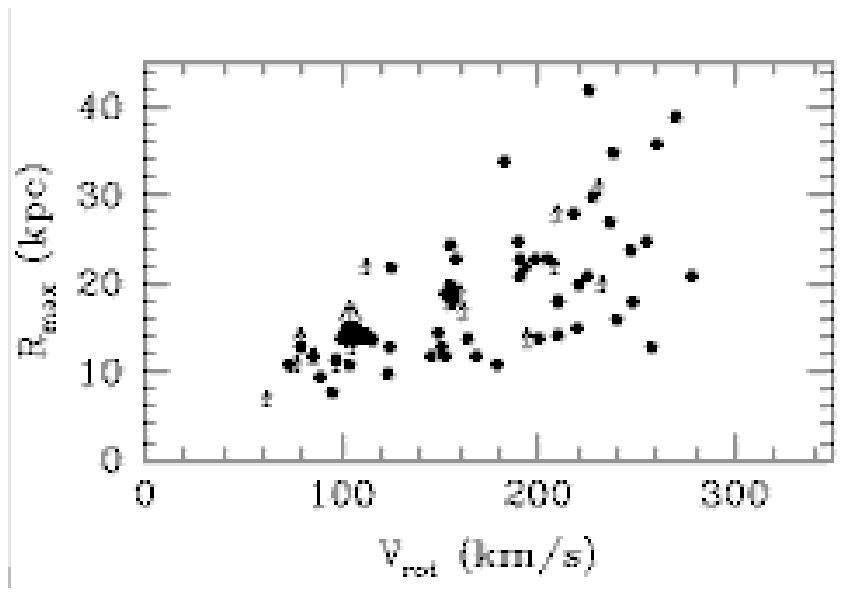}
\includegraphics[width=6.5cm]{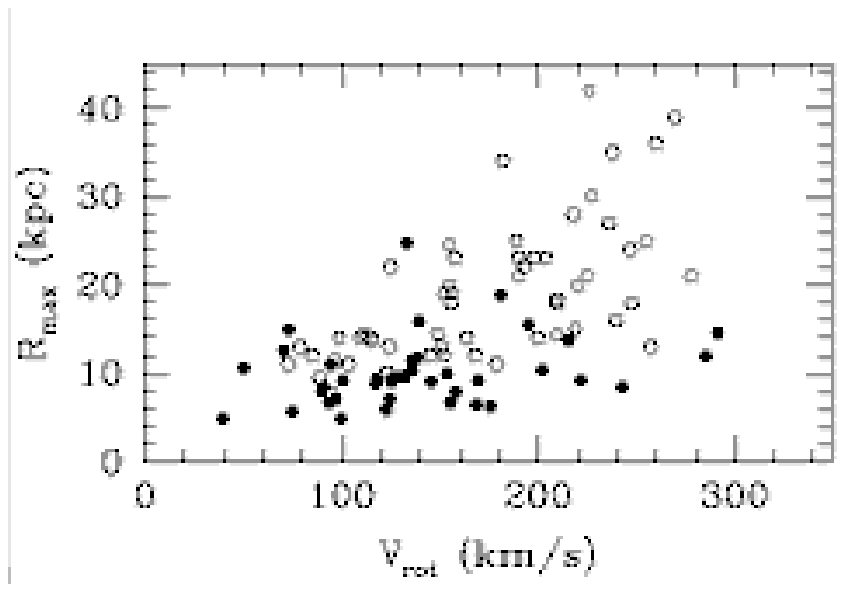}
\includegraphics[width=6.5cm]{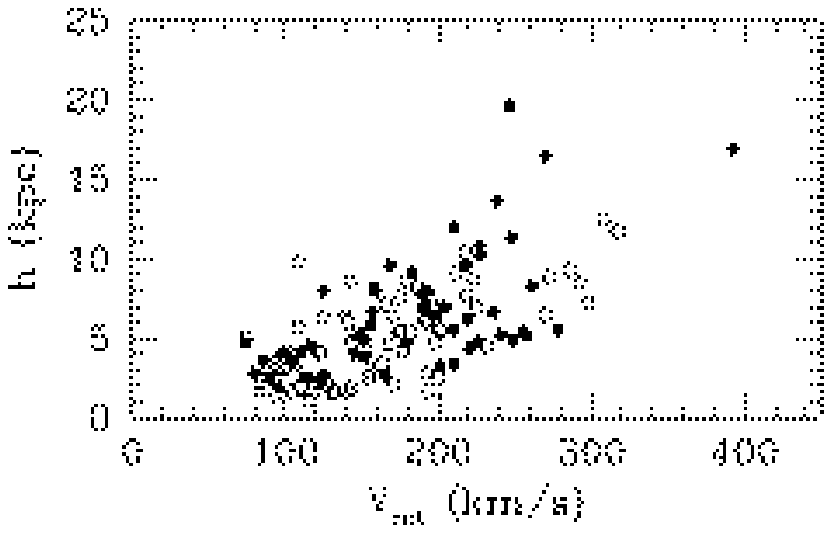}
\includegraphics[width=6.5cm]{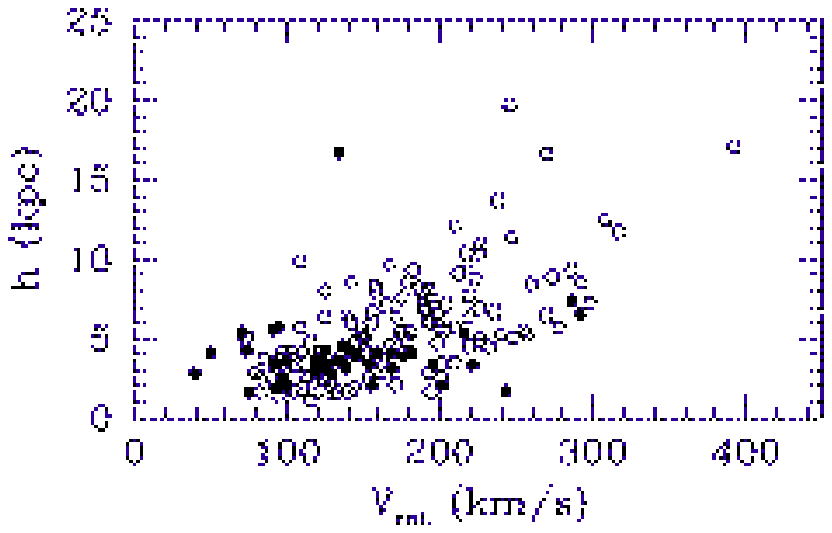}
\caption{Plots of $R_{\rm max}$ (top) and $h$ (bottom)
versus $V_{\rm rot}$. The top-left panel has all edge-on galaxies,
including lower limits.
The large dot plus lower limit is NGC 300. In the top-right panel  
the edge-on systems are open circles; the filled dots are moderately 
inclined galaxies of truncation Type II. 
The lower-left panel has all edge-on systems (dots) and
moderately inclined systems from \citet{deJ96a} (circles). 
The panel lower-right has all points from panel lower-left as open 
circles and those from moderately inclined systems op Type II as filled dots.}
\end{figure}
First we will look at the ratio $R_{\rm max}/h$ as a function of
$V_{\rm rot}$, both of which are distance independent.
In Fig. 3 (top-left) I plot the data from the edge-on samples in Table 1.
We see that these samples agree reasonably well among each other
with $R_{\rm max}/h$ ranging between about 2 and 5 and that there is no
strong dependence on $V_{\rm rot}$.
In Fig. 3 (top-right) I compare this distribution to that for
moderately inclined systems of Type II (``downbending break''). 
For these galaxies the
distribution is similar to that for edge-on galaxies. The lower part of 
the figure has $R_{\rm max}/h$ versus $h$; this shows that the 
face-on samples lack galaxies of large scalelength.

Next I look at the correlation between $R_{\rm max}$ and $V_{\rm rot}$.
Fig. 4 (top-left) shows the data from the edge-on samples.
I also include the lower limits from these samples.
Out of curiosity I add the point for NGC 300 \citep{Bla05} at $V_{\rm rot}
\sim 105$ km/s, which has no truncation even at 10 scalelengths from
the center. Inspite of that it is not
outside the distribution observed in edge-on systems. 
The reason is that it has a unusually
small $h$ for its $V_{\rm rot}$; {\it not an unusual extent} (in kpc) compared 
to its rotation.

In Fig. 4 (top-right) I have added the data for the moderately inclined
galaxies. We see that the two samples have different distributions,
the sample of Pohlen \&\ Trujillo showing smaller $R_{\rm max}$,
which may actually be expected from their method.
But it also shows an absense of correlation of $R_{\rm max}$ with
$V_{\rm rot}$, while these are clearly correlated in edge-on systems.
Further study is required to establish whether or not truncations found
in edge-on galaxies and in moderately inclined ones are indeed at the same 
thing.

Look now at $h$ versus $V_{\rm rot}$, which is a
well-known scaling relation for spirals.
In Fig. 4 (lower-left) the filled dots are the complete edge-on sample and the
open circles are the sample of moderately inclined galaxies from
\citet{deJ96a}. Note that the distributions are very
similar, suggestig that there is no systematic difference in scalelengths
measured in edge-on and face-on samples.
Next I present the de Jong points also as open circles (Fig. 4 - lower-right) 
and compare with the
Pohlen \&\ Trujillo sample of Type II (filled dots).
These points again do not show the expected correlation.

I conclude from this section as follows:
{\bf (1)} Measurements of truncations in edge-on and face-on systems
have fundamental differences.
{\bf (2)} There are no systematic differences in measurements
of the scalelength in edge-on and face-on samples.
{\bf (3)} It is not clear that in the study of truncations in face-on systems
by Pohlen \&\ Trujillo (2006) one is consistently deriving the same features
as in edge-on galaxies.
{\bf (4)} In particular the lack of a correlation of truncation radius
(and scalelength) with rotation velocity needs
clarification; is this due to sample peculiariteis or to the 
deprojection method?

\section{Truncations and warps}

Warps in the HI in external galaxies are most readily
observed in edge-on systems as NGC
5907 \citep{San76}. NGC 5907 has a clear and sharp truncation 
\citep{vdKS82} in its stellar disk at the same radius where
the warp starts. The ``prodigious warp'' in
NGC 4013 \citep{Bot87,Bot96} is very symmetric
and starts suddenly near the end of the optical disk. Inside the warp the 
stellar disk and HI-layer are exceedingly flat.
The stellar disk has a clear truncation \citep{vdKS82}. The
three-dimensional analysis by Bottema confirms
that in deprojection the warp starts very close to the
truncation radius.

NGC 628 is almost completely face-on.
The HI-velocity field displays a complicated pattern;
in the tilted-ring model the rings actually go through
the plane of the sky \citep{Sho84}. After subtraction of this
rotation field the residuals show that systematic vertical motions in the 
gas are less than a few km/s, indicating again that the disk is extremely flat. 
The radial luminosity profiles \citep{vdK88}
show evidence for a truncation, which again coincides with the onset of the
warp. 

So we see that, stellar disks and their accompanying HI-layers
are \underline{extremely}
flat (except near the edges
where sometimes minor optical warps are found) and
HI-warps often start \underline{suddenly}
at about the truncation radius of the stellar disk \citep[see also][]{vdK01}.

In his Ph.D. thesis Garc\'{\i}a-Ruiz \citep[see][]{Gar02}
presented HI observations of a sample of edge-on
galaxies. His sample consisted of 26 edge-on galaxies of
which at least 20 show evidence for an HI warp.
Unfortunately, the optical surface photometry
could not be calibrated. In a recent paper \citep{vdK07},
I investigated whether the Sloan Digital Sky Survey (SDSS) can be used to
see if there are truncations in this sample and if so, where the warps 
start with respect to these. In Fig. 5 I show two examples: 
UGC 7774 and UGC 6126 both have truncations;
in UGC 7774 the HI-warp starts on the sky at $R_{\rm max}$ while
in UGC 6126 as at a radius inside $R_{\rm max}$.
The observed distribution of the ratio $R_{\rm warp}/R_{\rm max}$
in the sample is statistically consistent with  that for
a random distribution of viewing angles in the situation,
where {\it all} warps start at about 1.1 $R_{\rm max}$. HI-warps
and truncations therefore seem related.

\begin{figure}[!ht]
\includegraphics[width=13cm]{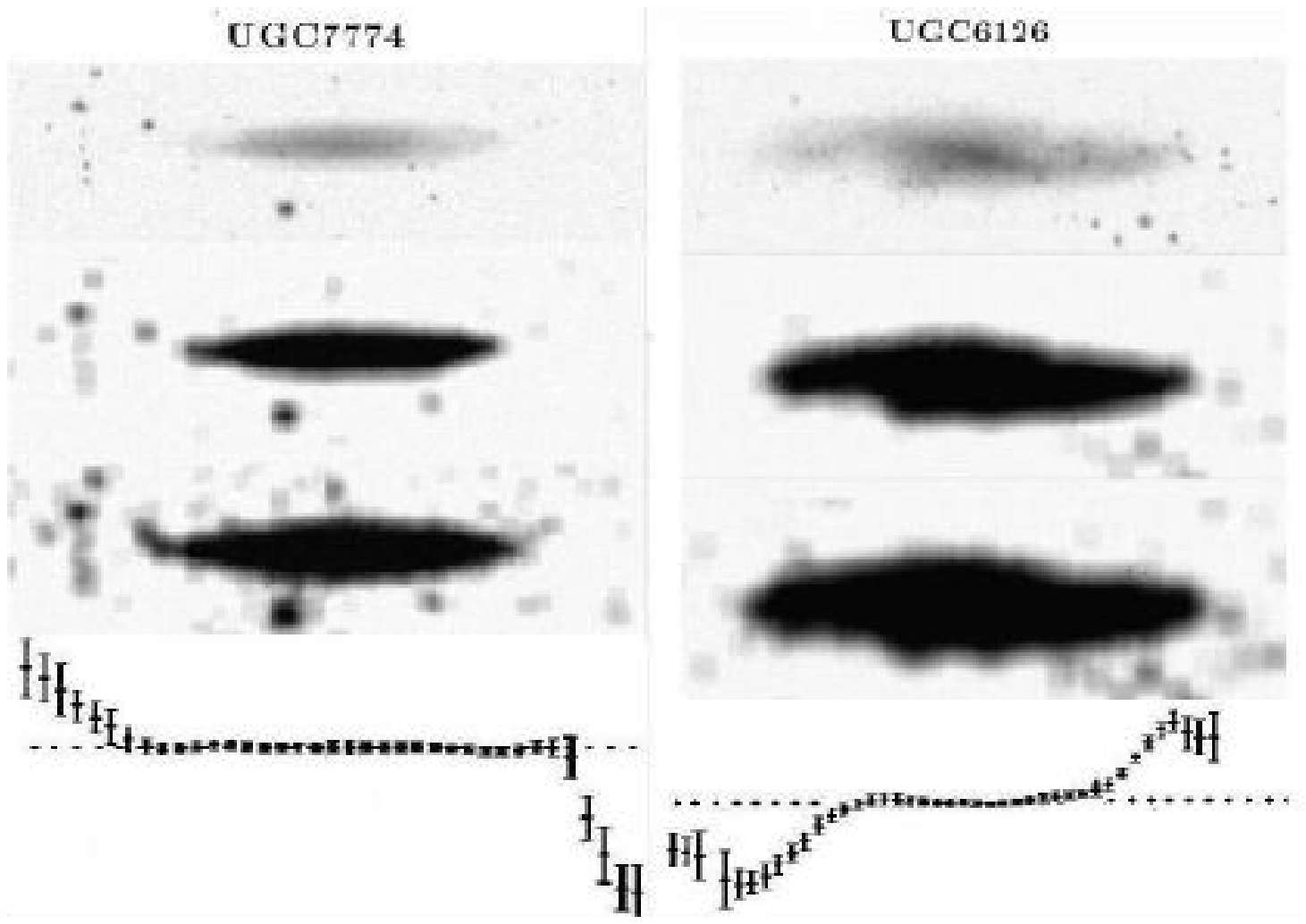}
\caption{UGC 7774 and UGC 6126 \citep[from][]{vdK07}.}
\end{figure}

The {\bf properties of HI-warps} have been described by \citet{Bri90}
in the form of a set of rules of behaviour; with subsequent information we 
can summarize the most important properties as follows (for a more extended
discussion see van der Kruit, 2007):\\
$\ast $ All galaxies with HI extending beyond their optical disk have warps
\citep{Gar02}.\\
$\ast $ Many galaxies have relatively sharp truncations in their stellar 
disks.\\
$\ast $ In edge-on galaxies the HI warps sets in just beyond
the truncation radius,  for less inclined systems it sets
in near the Holmberg radius.\\
$\ast $ In many cases the rotation curve shows a feature that indicates that
there is at the truncation radius also a sharp drop in mass surface density.\\
$\ast $ The onset of the warp is abrupt and discontinuous
and there is a steep slope in HI-surface density at this point.\\
$\ast $ Inner disks are extremely flat (with at most
minor optical warps) and the HI-warps define a single ``new reference frame''
\citep{Bri90}.

This leads to the following features of a possible formation scenario
\citep{vdK87,vdK01,vdK07}:
{\bf (1)} The inner disk (mostly stars) and the warped outer disk (mostly HI)
are distinct components.
{\bf (2)} These probably have distinct formation histories
and formed during different epochs.
{\bf (3)} Inner disks form initially and settle as
rigid, flat structures with well-defined boundaries (truncations)
corresponding to a maximum in the specific angular momentum distribution.
{\bf (4)} If the initial distribution resembles a Mestel sphere and if the
collapse occurs with approximate conservation of specific angular momentum
truncated, exponential disks result automatically.
{\bf (5)} HI-warps result from later infall of gas with a different
orientation of angular momentum.
{\bf (6)} The often regular structure of the warps and 
Brigg's ``new reference frame''
may result from re-arranging the structure from individual infalling gas
clouds by interactions with neighbours or with
an intergalactic medium.

\section{Conclusions}

$\bullet $ Truncations are a common feature in edge-on stellar disks.\\
$\bullet $ The relation of truncations as observed in moderately inclined 
systems to those in edge-on galaxies needs further clarification, in
particular the absence of a correlation of $R_{\rm max}$ with $V_{\rm rot}$
and $h$ in the face-on sample.\\
$\bullet $ The origin of truncations is most likely related to a maximum
in the specific angular momentum in the material that formed the stellar
disks, involving modest redistribution of angular momentum.\\
$\bullet $ Stellar disks and their accompanying gas-layers are extremely 
flat.\\
$\bullet $ HI-warps start just beyond the truncation radius and 
stellar disks and HI-warps appear to be distinct components.\\
$\bullet $ This suggests that inner disks form initially and 
settle as rigid, very flat structures, while HI-warps result from later infall
of gas with a different orientation of angular momentum.



\end{document}